\documentclass[aip,jap,preprint]{revtex4-1}
\pdfoutput=1
\draft 
\usepackage{graphicx}%
\usepackage{dcolumn}%
\usepackage{bm}%
\usepackage{natbib}

\begin{document}


\title{Anisotropic Friedel oscillations inside the domain wall} 



\author{R. Ghanbary}
\affiliation{Department of Physics, Payame Noor University,Urmia}
\author{A. Phirouznia}
\affiliation{Department of Physics, Azarbaijan Shahid Madani
University, 53714-161, Tabriz, Iran\\ \& Condensed Matter Computational Research Lab. Azarbaijan Shahid
Madani University, 53714-161, Tabriz, Iran}


\date{\today}

\begin{abstract}
The influence of the non-collinear magnetic configuration on Friedel is investigated theoretically. Specifically the influence of the magnetic configuration on the induced electric charge in a N\'{e}el type domain wall (DW) has been obtained. The well-known Levy and Zhang eigenstates for a linear DW have been employed. Then the dielectric function of this magnetic system has been obtained within the random phase approximation. Results of the current work demonstrated that magnetic configuration of the system manifests itself in the electric properties such as induced charge distribution. Meanwhile the anisotropy of the induced charge distribution in the real space provides a measurable way for the determination of the DW orientation. In addition anisotropy of the dielectric function in k-space arises as a result of the anisotropy of the magnetic configuration. Therefore the orientation of the magnetic DW could also be captured by full optical measurements.
\end{abstract}


\maketitle 

\section{Introduction}
An extremely big effort has been devoted to explore and the
realization of the quantum nature of matter in the field of
condensed matter physics \cite{STM,zwerger,aronov}. According to the historical point of view
it was generally accepted that this developing field could emerge
many exciting phenomena,  which has made this research field very
active for applications as well as fundamental studies \cite{kragh}.
\\
Meanwhile, a fundamental understanding of the system properties
requires a deep knowledge of quantum many particle effects in the
system \cite{friedel1953,Kittel,bruno,STM,aronov}. Soon after the development of quantum based approaches, it
was realized that dynamical dielectric function, captures some of
the most important many particle induced effects of any given
system \cite{horing}. For example dynamical dielectric function determines the
screening of the coulomb interaction, optical absorption spectra,
Friedel oscillations and plasmon dispersion relation of the system \cite{Wunsch,Pyatkovskiy}.
\\
Friedel oscillation is an obvious example of the quantum nature of
the materials \cite{friedel1953} which manifests itself in the density redistribution
response to an external charged impurity \cite{Sarma,badalyan,Giuliani}. Despite the similarities
between the different Friedel oscillations in two-dimensional
electron gas systems, there is also a  fundamental deviation
depending on the system form factor identified by the Hamiltonian
eigenstates which results in a anisotropic Friedel oscillations in
particular cases that deserves some attention. In this context it is
important to know that such an anisotropic Friedel oscillation  can
be achieved in magnetic modulated structures. It was also realized that 
the anisotropic Friedel oscillations could be generated when there is a major contribution from wave numbers which do not
correspond to Fermi wave vector \cite{hofmann}.
\\
In the present work we have shown that Friedel oscillations of a
two-dimensional domain wall (DW) are highly anisotropic in which the
spatial distribution of induced charge density, extremely depends on
the orientation of the position vector relative to the DW
characteristic vectors. These characteristic vectors which determine
the spatial configuration of the DW are the DW
rotation axis, $\hat{n}(\textbf{r})$ and a unit vector defined along the DW width.  This
anisotropy seems to be quite reasonable due to the magnetic
anisotropy, induced in the presence of the DW. This could
be considered as an example of the dependence of electric properties
on magnetic configuration in the system.
\\
Another pronounced example in this
field is the influence of the DW on electric resistivity
i.e. The magnetoresistance of the system. The effect of the DW on the magnetoresistance has indeed been a controversial
subject in the field of spin dependent transport. It was realized
that the DW can either increase or decrease the electrical
resistivity \cite{Greg,Lep,Levy,Rudi,cetin,Tatara}.
\\
The anisotropic behavior of the Friedel oscillations
could be captured quite well within the random phase approximation.
Since the Friedel oscillations in the real space can be detected experimentally\cite{imaging1,imaging2,imaging3}. 
Understanding the dependence of the induced charge distribution on
the magnetic order, could help us to development of new approaches
for better experimental determination of the DW orientation
and the configuration of the DW bulging by Friedel oscillation measurements. 
\\
The primary focus of the present work was on
showing the effect of the magnetic configuration of the system on
dielectric function and Friedel oscillations of a two-dimensional
system. In the most of the previous works Friedel oscillations show
a spherical symmetry without any directional dependence of the
induced charge density. This is due to the fact that in these cases
the system actually contains the spherical or circular symmetry which manifests itself in the
Friedel oscillations or the anisotropy of the band structures in
k-space might be significant only at especial ranges of relatively high
energies, containing non-contributing levels.
\section{Model and approach}
In the current study we have considered the linear the smooth DW between the oppositely directed ferromagnetic regions. In this case the Hamiltonian of the system reads
\begin{eqnarray}
H=\frac{{{p}^{2}}}{2m^*}+J\hat{\sigma }.\hat{m}(x),
\end{eqnarray}
in which $m^*$ is the effective mass of the electrons in the periodic lattice, $J$ denotes the
exchange interaction strength, $\sigma$ stands for the spin
operator in terms of the Pauli spin matrices,
$\hat{m}(x)$ is the
unit vector directed along the local magnetization.
\\
The local direction of the magnetization has been assumed to vary along the x-axis as $\theta(x)=(\pi/d)x$ in which $\theta $ characterizes the direction of the magnetic moments in the system plane and $d$ is the DW width. Then within the Levy and Zhang perturbative approach the eigen-states of the linear domain walls (DWs) are given by
\begin{eqnarray}
{{\psi }_{\uparrow }}(\vec{k},\vec{r})&=&\frac{\alpha
({{k}_{x}})}{\sqrt{{{L}_{x}}}{{L}_{z}}}\left[ {{R}_{\theta }}\left(
\begin{array}{cc}
   {{e}^{i{{{\vec{k}}}_{\uparrow }}.\vec{r}}} \\
  0 \\
\end{array} \right)-i\frac{{{k}_{x}}}{{{k}_{F}}}\xi {{R}_{\theta }}
\left( \begin{array}{cc}
   0 \\
  {{e}^{i{{{\vec{k}}}_{\uparrow }}.\vec{r}}} \\
\end{array} \right) \right]
\nonumber\\
{{\psi }_{\downarrow }}(\vec{k},\vec{r})&=&\frac{\alpha
({{k}_{x}})}{\sqrt{{{L}_{x}}}{{L}_{z}}}\left[ {{R}_{\theta }}\left(
\begin{array}{cc}
  & 0 \\
 & {{e}^{i{{{\vec{k}}}_{\downarrow }}.\vec{r}}} \\
\end{array} \right)-i\frac{{{k}_{x}}}{{{k}_{F}}}\xi {{R}_{\theta }}\left( \begin{array}{cc}
  & {{e}^{i{{{\vec{k}}}_{\downarrow }}.\vec{r}}} \nonumber\\
 & 0 \\
\end{array} \right) \right].\\
\end{eqnarray}
Where,
${\alpha}(k_x)=(2(1+k_{x}^2\xi^{2}))^{-\frac{1}{2}}$ is the
normalization factor, $\xi=\hbar^2\pi^2 k_F/(4m^* J
d^2)$ is a dimensionless parameter which measures the strength of the exchange energy.
$R_{\theta}=exp(-i(\theta(x)/2)\hat{\sigma}\cdot\hat{n})$ is the spin
rotation operator, $\hat{n}$ is the direction
of the DW rotation axis in which for the current system containing two-dimensional N\'{e}el type DW, $\hat{n}$ should be directed along the plane normal axis and . In addition the system eigen-values are as follows
\begin{eqnarray}
{{\epsilon }}_{ {{k}_{\sigma }}\sigma}=\frac{{{\hbar
}^{2}}k_{\sigma }^{2}}{2{{m}^{*}}}+\sigma J+\xi J,
\end{eqnarray}
in which $\sigma=+(-)$ corresponds to the ${\psi }_{\uparrow }$ $({\psi }_{\downarrow })$ eigen-state.
The the non-interacting Lindhard response function is then given by
\begin{eqnarray}
{{\Pi }_{0}}(\vec{q},\omega
)=\frac{1}{S}\,\sum\limits_{\vec{k}\sigma
}{\frac{{{n}_{\vec{k}\sigma }}-{{n}_{\vec{k}+\vec{q}\,{\sigma
}'}}}{\hbar \omega +{{\epsilon }_{\vec{k}\sigma }}-{{\epsilon
}_{\vec{k}+\vec{q}\,{\sigma }'}}+i\eta}}\,{{F}_{\sigma {\sigma
}'}}(\vec{k},\vec{k}+\vec{q}). \nonumber\\
\end{eqnarray}
In which $S$ denotes the area of the system, ${n}_{\vec{k}\sigma}$ is the state occupation number given by the Fermi distribution, $\eta$ is a small positive number and the form factor ${{F}_{\sigma {\sigma}'}}(\vec{k},\vec{k}+\vec{q})$ measures
the overlap of spinor wave functions at different k-points. This form factor can be expressed by the following relation
\begin{eqnarray}
{{F}_{\sigma {\sigma }'}}(\vec{k},\vec{{k}'})={{{\left|
\,\left\langle  {{\psi }_{{k}'{\sigma }'}}
\right|{{e}^{-i\,\vec{q}.\vec{r}}}\left| {{\psi }_{k\sigma }}
\right\rangle  \right|}^{2}}}.
\end{eqnarray}
Accordingly the form factor of different spin-bands are given as
\begin{eqnarray}
\label{form1}
{{F}_{\downarrow \uparrow
}}(\vec{k},\vec{k}+\vec{q})&=&{{F}_{\uparrow \downarrow
}}(\vec{k},\vec{k}+\vec{q})\\&=&{{\alpha }^{4}}{{\xi
}^{2}}{{q}^{2}}{{\cos }^{2}}({{\theta }_{q}})\nonumber\\
\label{form2}
{{F}_{\uparrow \uparrow
}}(\vec{k},\vec{k}+\vec{q})&=&{{F}_{\downarrow \downarrow
}}(\vec{k},\vec{k}+\vec{q})\\&=&{{\alpha }^{4}}(1+\xi^{2}kq\cos\theta_q\cos\theta_k+\xi^{2}k^2\cos^2\theta_k)^{2}\nonumber
\end{eqnarray}
The real and imaginary parts of the response function are given by
\begin{eqnarray}
\textit{Re}~{{\Pi }_{0}}(\vec{q},\omega
)&=&\frac{1}{{{S}}}\,P\sum\limits_{\vec{k}\sigma
}{\frac{{{n}_{\vec{k}\sigma }}-{{n}_{\vec{k}+\vec{q}\,{\sigma
}'}}}{\hbar \omega +{{\varepsilon }_{\vec{k}\sigma }}-{{\varepsilon
}_{\vec{k}+\vec{q}\,{\sigma }'}}}}\,{{F}_{\sigma {\sigma
}'}}(\vec{k},\vec{k}+\vec{q})\nonumber\\
{\textit{Im}}~{{\Pi }_{0}}(\vec{q},\omega )&=&-\frac{\pi
}{S}\,\sum\limits_{\vec{{k}'}{\sigma }'}{\sum\limits_{\vec{k}\sigma
}}{{F}_{\sigma {\sigma }'}}(\vec{k},\vec{{k}'})\\&&~~~~~~\times{({{n}_{\vec{k}\sigma
}}-{{n}_{\vec{{k}'}{\sigma }'}})
\delta (\hbar \omega +{{\epsilon
}}_{\vec{k}\sigma }-{{\epsilon }}_{\vec{{k}'}{\sigma }'})}.\nonumber
\end{eqnarray}

Finally within the random phase approximation (RPA) response function and dielectric function of the system are given by the following expressions respectively
\begin{eqnarray}
\Pi^{RPA}(\vec{q},\omega )=\frac{{{\Pi
}_{0}}(\vec{q},\omega )}{1-{{v}_{q}}{{\Pi }_{0}}(\vec{q},\omega )}\\
{{\varepsilon }^{RPA}}(\vec{q},\omega )=1-{{v}_{q}}{{\Pi
}_{0}}(\vec{q},\omega ).
\end{eqnarray}
In which ${v}_{q}=\frac{2\pi e^2}{q}$ is the k-space Coulomb interaction of a two-dimensional system.
\\
Finally the density of the induced charge by a single charged impurity is given as
\begin{eqnarray}
{{\delta n }_{ind}}(\vec{r})=\frac{1}{{{(2\pi
)}^{3}}}\int{{{v}_{q\,}}\left[ \frac{{{\Pi }_{0}}(\vec{q},\omega
)}{1-{{v}_{q}}{{\Pi }_{0}}(\vec{q},\omega )}
\right]}\,\,{{e}^{i\,\vec{q}.\vec{r}}}d\vec{q}.
\end{eqnarray}
The oscillations of this charge distribution known as the Friedel oscillations.
\section{Results and discussions}
\begin{figure}
\includegraphics[width=8.0cm]{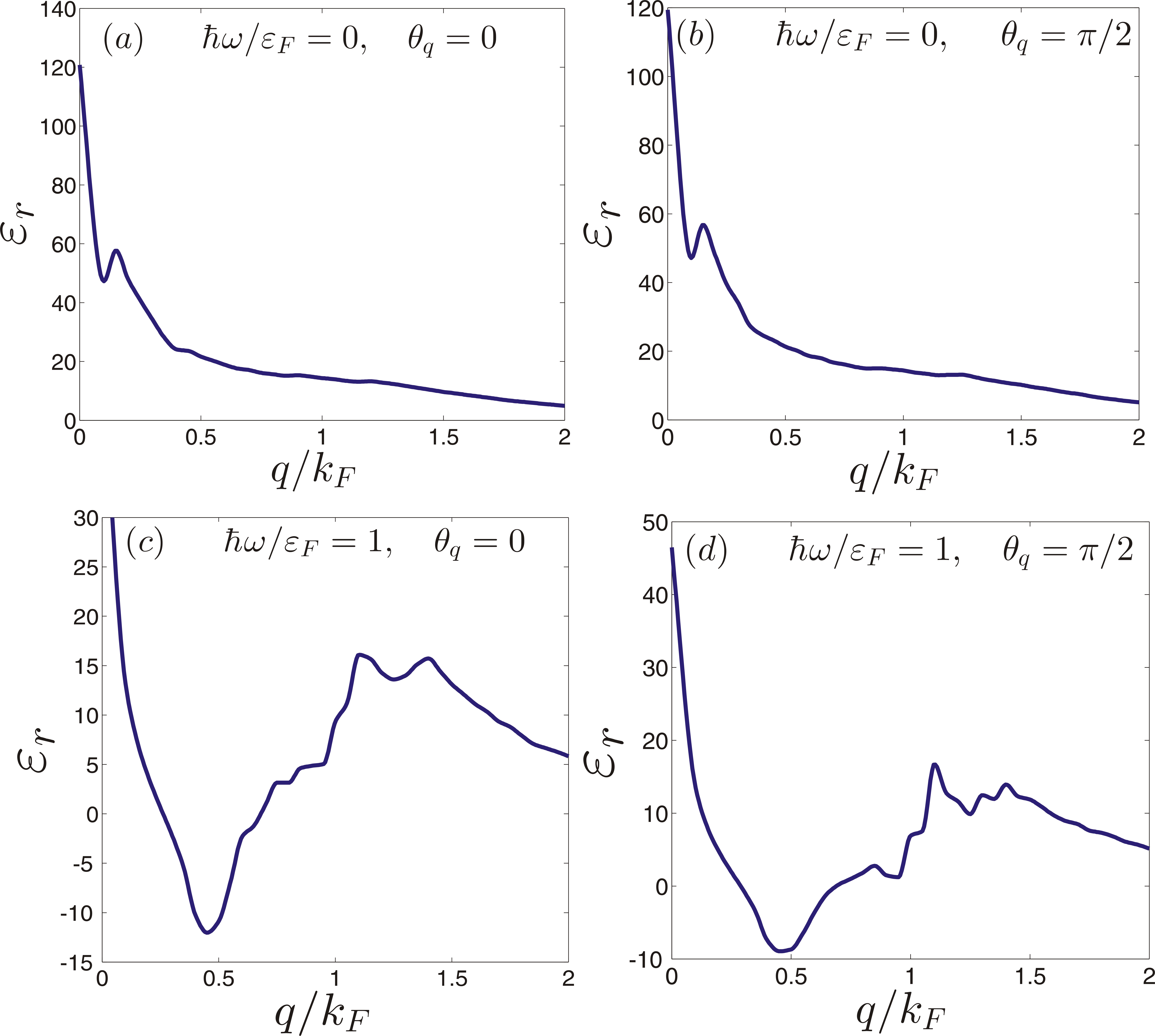}%
\caption{Real part of the dielectric function in different directions of the k-space identified by $\theta_q$. The exchange coupling constant has been assumed to be $J=0.3$eV and $\xi=0.5$.\label{epr}}%
\end{figure}

\begin{figure}
\includegraphics[width=9.2cm]{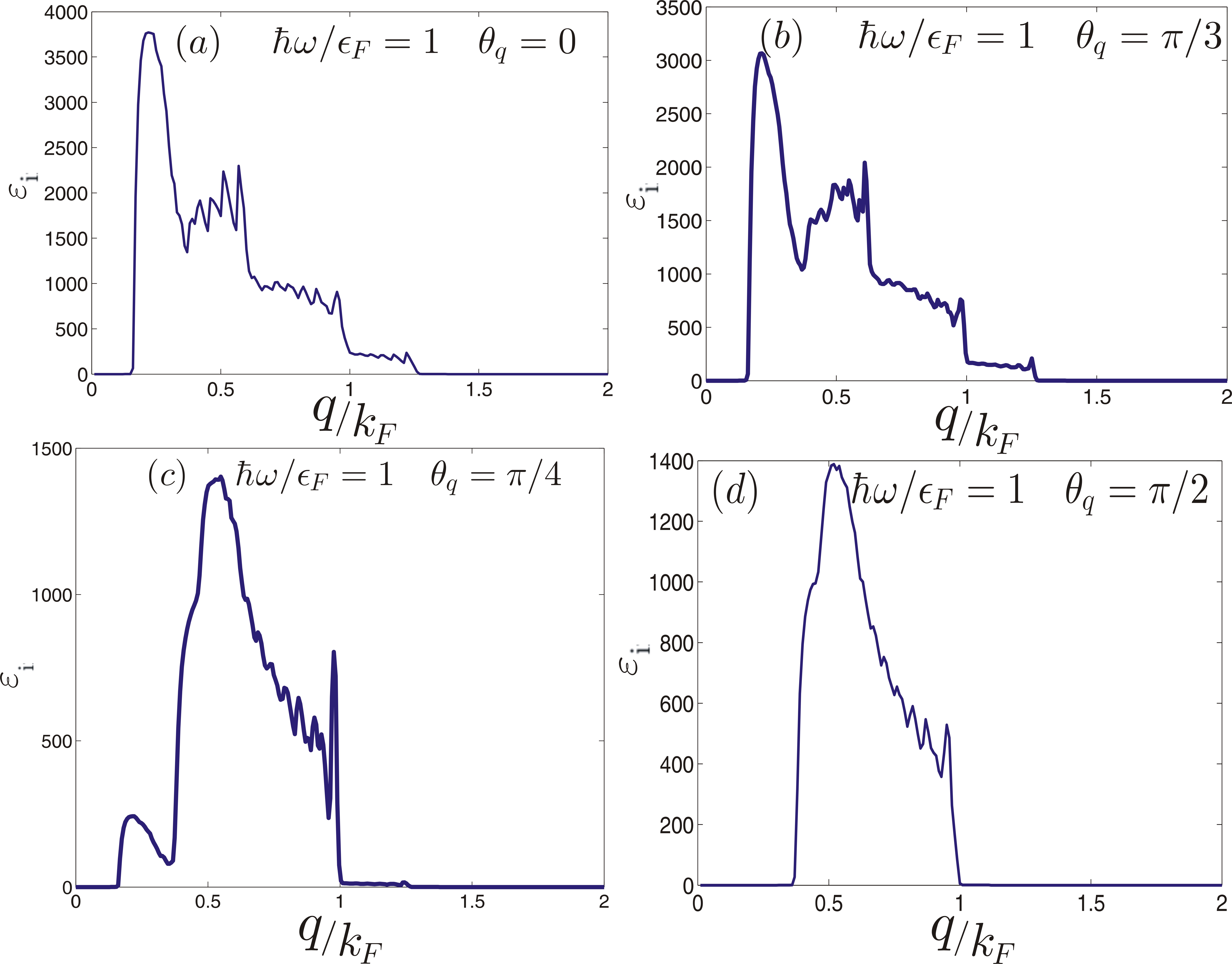}%
\caption{Imaginary part of the dielectric function in different directions of the k-space identified by $\theta_q$. The exchange coupling constant has been assumed to be $J=0.3$eV and $\xi=0.5$.\label{epi}}%
\end{figure}

\begin{figure}
\includegraphics[width=7.5cm]{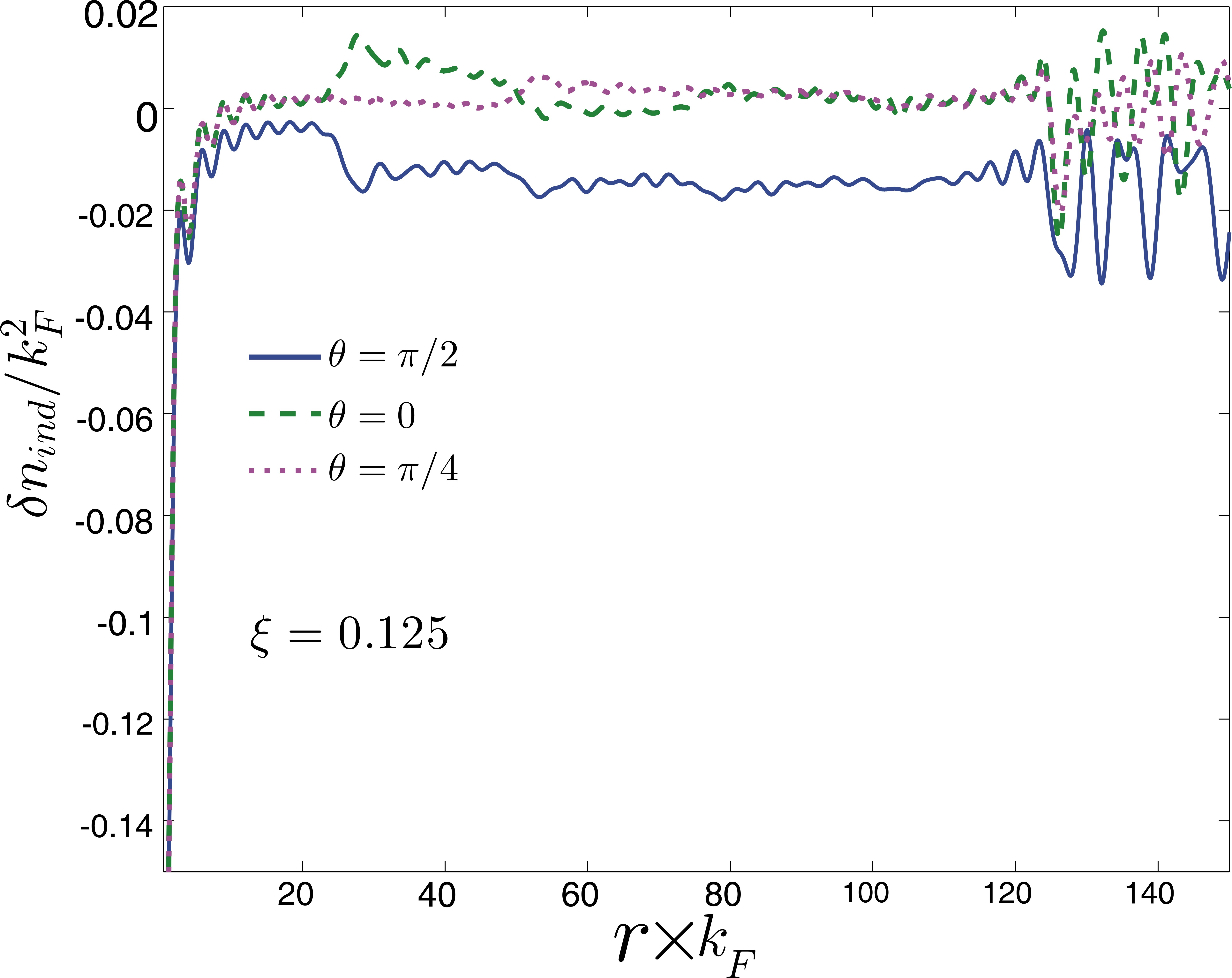}%
\caption{Anisotropic Friedel oscillations at different directions. The impurity has been placed in the origin and the exchange interaction strength has been assumed to be $J=0.2$eV. \label{2d}}%
\end{figure}

\begin{figure}
\includegraphics[width=8.5cm]{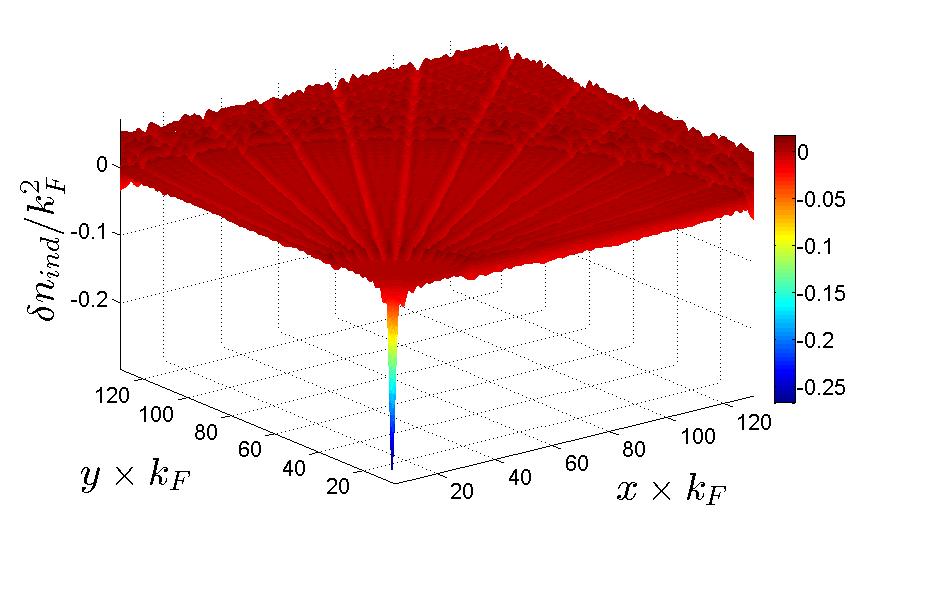}%
\caption{Anisotropic Friedel oscillations at different directions. The impurity has been placed in the origin ($r=0$) and the exchange interaction strength has been assumed to be $J=0.3$eV where we have choose $\xi=0.083$. As shown in this figure at some specific directions the density of the induced charge takes takes quite different values. Meanwhile at the other directions the density is nearly the same. \label{3d}}%
\end{figure}

We have used parameters of typical parameters by taking $k_F=5.1\times
{{10}^{+9}}\,{{m}^{-1}}$, $T=300$K and $\epsilon_F=1$eV. We have considered in N\'{e}el type magnetic DW in which the direction of the magnetic station varies slowly along the domain. Size of the DW has been considered as long as possible within a typical range of the smooth DWs. In this work we have considered DW width to be $d=100$nm. This choice of slowly varying DWs is due to the fact that period of the Friedel oscillations should be much smaller than the size of the magnetic DW. On the other hand to perturbative approach which has been employed to obtain the DW eigen-states has been totally fall down for a non-linear sharp DW in which the local direction of the magnetization changes abruptly.
\\
As discussed before we have obtained dielectric function and real space Friedel oscillations within a smooth DW. We have employed RPA approximation and typical DW parameters.
\\
In Figs. \ref{epr} and \ref{epi} real and imaginary parts of the DW dielectric function has been depicted as a function of the wave number, $q$, for several values of the k-space polar angle, $\theta_q=\arctan(q_y/q_x)$, at different frequencies ($\omega$).
\\
As shown in Fig. \ref{epr} real part of the dielectric function has a strongly modulated by the photon energy and the orientation of the wave number could be considered as a minor factor in the real dielectric function. However numerical results show that the dependence of the real dielectric function on photon energy could be effectively change by magnitude of the wave number.
These two parameters (i.e. the photon energy and the wave number) could not be treated on the same footing. Since as shown in this figure the photon energy could result in a order of magnitude change of the real dielectric function (Fig. \ref{epr}). Meanwhile at a given photon energy dielectric function shows an anisotropic functionality in k-space.
\\
The above discussion holds for imaginary part of the dielectric function as shown in Fig. \ref{epi}. Imaginary part of the dielectric function measures external field absorption. Fig. \ref{epi} shows anisotropic features of the absorption spectra. As shown in this figure imaginary part of the dielectric function changes for different orientations of the wave number. This means that the energy absorption effectively depends on the direction of the photon wave number and therefore the photon polarization. For different directions of the polarization different values of the energy absorption is obtained. Therefore this anisotropic absorption effect could be utilized in the detection of DW orientation by a full optical measurements. This means that the analysis of the directional dependence of the absorption spectra reveals the DW orientation.
\\
Meanwhile it should be considered that the anisotropic effects can also be appeared when the imaginary part of the dielectric function is depicted as a function of the photon energy, $\hbar\omega$ at given wave numbers. In this case the effective absorption range increases by the transferred momentum. Increasing the photon wave vector shifts the effective range of the absorption curve to the high energies. This can be explained as a trivial result if we consider that the transferred momentum ($q$) increases by the photon frequency.
\\
\\
The effect of the single impurity in the charge density variation is determined by the static density response function. Results of the current work show that the magnetic structure of the system directly manifests itself in the induced charge of distribution of the system. As shown in Figs. \ref{2d} and \ref{3d} the impunity induced charge inside the DW has an anisotropic distribution inside the system. This charge distribution inside the DW is significantly depends on the direction the investigation. As shown in Fig. \ref{2d} the distribution of the induced charge is nearly the same for some of the directions in the real space i.e. for $\theta=0$ and $\theta=\pi/2$, however even the sign of the induced charge is different in the $\theta=\pi/4$ direction. At first look one might think about the isotropic band energies in k-space which have mentioned as ${{\epsilon }}_{ {{k}_{\sigma }}\sigma}=\frac{{{\hbar
}^{2}}k_{\sigma }^{2}}{2{{m}^{*}}}+\sigma J+\xi J$. Therefore isotropic induced charge distribution in real space could be expected. However, anisotropy of the spin-dependent form factors in k-space (Eqs. \ref{form1} and \ref{form2}) results in the anisotropy of the k-space dielectric function and also the real space induced charge.
\\
In collinear magnetic systems where we have $F_{\sigma \sigma'}\sim \delta_{\sigma \sigma'}$ the form factor of the system cannot capture a detailed information about the magnetic configuration of the system and therefore we cannot make any judgment about the magnetic configuration of the system. Meanwhile as mention in the present case of the non-collinear magnetic systems, full optical observations could result in  detection of the magnetic orientation of the DW.
\\
In addition it can be inferred that in the presence of the impurities DW shows different values of the electric resistance for different geometries. This means that the electric resistance is different for current in wall (CIW) and current perpendicular to wall (CPW) geometries. Since the contribution of the charged impurities in the resistivity could be considered from two point of view. First the scattering of the moving carriers which has been made by the impurity itself. The second contribution come from the scattering of the moving electrons by the impurity induced charged distribution. Since the last contribution originates from the anisotropic charge distribution therefore the resistivity of the CIW and CPW geometries would be quite different. It should be noted that this the contribution of the induced charge distribution in the resistivity could be taken into account when the electron-electron interactions have not been ignored in the given approach.
\\
Transport phenomenon is actually a nonlocal process in the quantum theory of transport. Since a moving particle in the quantum theory of transport has been considered as a problem of wave packet propagation in which the dynamics of the wave packet has been the determined by all of the interactions which have been taken place in the whole area of the DW. Therefore regardless of the local effects that could be made by a single short-range impunity  since the long-range induced charge density could also contribute in the propagation of the wave packet therefore one can expect anisotropic transport effects as a result of the anisotropy of the induced charge distribution.
\\
\\
\section{Concluding remarks}
Result of the present work show that the non-collinear magnetic structures such as DWs have a measurable effects in the anisotropy of the dielectric function and impurity induced charge distribution.
Results of the current study could provide an experimental framework for the detection of the DW configuration. Since the anisotropy of the dielectric function could be determined by the optical absorption measurements. Meanwhile detection of the charge of distribution experimentally available at the present time. Electric measurements can be employed for the direction of the mentioned anisotropic charge distribution and therefore the direction of the DW orientation inside the system could be obtained. It is interesting to know that generally the DW is pinned by a single impurity. A moving DW can be stopped and captured by an impurity. Therefore it is expected that a DW at rest would contain impunity.
\\


%
%

%


\bibliography{ref}
\bibliographystyle{unsrtnat}

\end{document}